\documentclass[slac_one]{revtex4}
\usepackage{graphicx}
\usepackage{fancyhdr}

\begin{document}

\title{Kaon experiments at CERN: recent results and prospects}

\author{Evgueni Goudzovski \small ~ for the NA48/2 and NA62 collaborations}
\affiliation{School of Physics and Astronomy, University of Birmingham, B15 2TT, United Kingdom}

\begin{abstract}
Recent results from the NA48/2 and NA62 kaon decay-in-flight experiments at CERN are presented. A precision measurement of the helicity-suppressed ratio $R_K$ of the $K^\pm\to e^\pm\nu$ and $K^\pm\to\mu^\pm\nu$ decay rates has been performed using the full dedicated data set collected by the NA62 experiment ($R_K$ phase); the result is in agreement with the Standard Model expectation. New measurements of the $K^\pm\to\pi^\pm\gamma\gamma$ decay at the NA48/2 and NA62 experiments provide further tests of the Chiral Perturbation Theory. A planned measurement of the branching ratio of the ultra-rare $K^+\to\pi^+\nu\bar\nu$ decay at 10\% precision is expected to represent a powerful test of the Standard Model.
\end{abstract}

\maketitle

\thispagestyle{fancy}

\section*{INTRODUCTION}

In 2003--04, the NA48/2 experiment has collected at the CERN SPS the world largest sample of charged kaon decays, with the main goal of searching for direct CP violation in the $K^\pm\to3\pi$ decays~\cite{ba07}. In 2007--08, the NA62 experiment ($R_K$ phase) collected a large minimum bias data sample with the same detector but modified data taking conditions, with the main goal of measuring the ratio of the rates of the $K^\pm\to\ell^\pm\nu$ decays ($\ell=e,\mu$). The large statistics accumulated by both experiments has allowed the studies of a range of rare $K^\pm$ decay modes. The main stage of the NA62 experiment, expected to start physics data taking in 2014, aims at  measuring the $K^+\to\pi^+\nu\bar\nu$ decay rate. The recent results and prospects of these experiments are discussed here.

\section{BEAM AND DETECTOR IN 2003--08}

The beam line has been designed to deliver simultaneous narrow momentum band $K^+$ and $K^-$ beams derived from the primary 400 GeV/$c$ protons extracted
from the CERN SPS. Central beam momenta of 60 GeV/$c$ and 74 GeV/$c$ have been used. The beam kaons decayed in a fiducial decay volume contained in a 114~m long cylindrical vacuum tank. A detailed description of the detector used in 2003--08 is available in~\cite{fa07}. The momenta of charged decay products are measured in a magnetic spectrometer, housed in a tank filled with helium placed after the decay volume. The spectrometer comprises four drift chambers (DCHs), two upstream and two downstream of a dipole magnet which gives a horizontal transverse momentum kick of $120~\mathrm{MeV}/c$ or $265~\mathrm{MeV}/c$ to singly-charged particles. Each DCH is composed of eight planes of sense wires. A plastic scintillator hodoscope (HOD) producing fast trigger signals and providing precise time measurements of charged particles is placed after the spectrometer. A 127~cm thick liquid krypton (LKr) electromagnetic calorimeter located further downstream is used for lepton identification and as
a photon veto detector. Its 13248 readout cells have a transverse size of approximately 2$\times$2 cm$^2$ each, without longitudinal segmentation.

\section{LEPTON UNIVERSALITY TEST WITH 2007--08 DATA}

Decays of pseudoscalar mesons to light leptons ($P^\pm\to\ell^\pm\nu$, denoted $P_{\ell 2}$ below) are suppressed in the Standard Model (SM) by helicity considerations. Ratios of leptonic decay rates of the same meson can be computed very precisely: in particular, the SM prediction for the ratio
$R_K=\Gamma(K_{e2})/\Gamma(K_{\mu 2})$ is~\cite{ci07}
\begin{equation}
\label{Rdef} R_K^\mathrm{SM} = \left(\frac{m_e}{m_\mu}\right)^2
\left(\frac{m_K^2-m_e^2}{m_K^2-m_\mu^2}\right)^2 (1 + \delta
R_{\mathrm{QED}})=(2.477 \pm 0.001)\times 10^{-5},
\end{equation}
where $\delta R_{\mathrm{QED}}=(-3.79\pm0.04)\%$ is an
electromagnetic correction. Within extensions of the SM involving two Higgs doublets, $R_K$ is sensitive to lepton flavour violating effects induced by loop processes with the charged Higgs boson ($H^\pm$) exchange~\cite{ma06}. A recent study~\cite{gi12} has concluded that $R_K$ can be enhanced by ${\cal O}(1\%)$ within the Minimal Supersymmetric Standard Model. However, the potential new physics effects are constrained by other observables such as $B_s\to\mu^+\mu^-$ and $B_u\to\tau\nu$ decay rates~\cite{fo12}. On the other hand, $R_K$ is sensitive to the neutrino mixing parameters within SM extensions involving a fourth generation of quarks and leptons~\cite{la10}.

\begin{figure}[tb]
\begin{center}
\resizebox{0.4\textwidth}{!}{\includegraphics{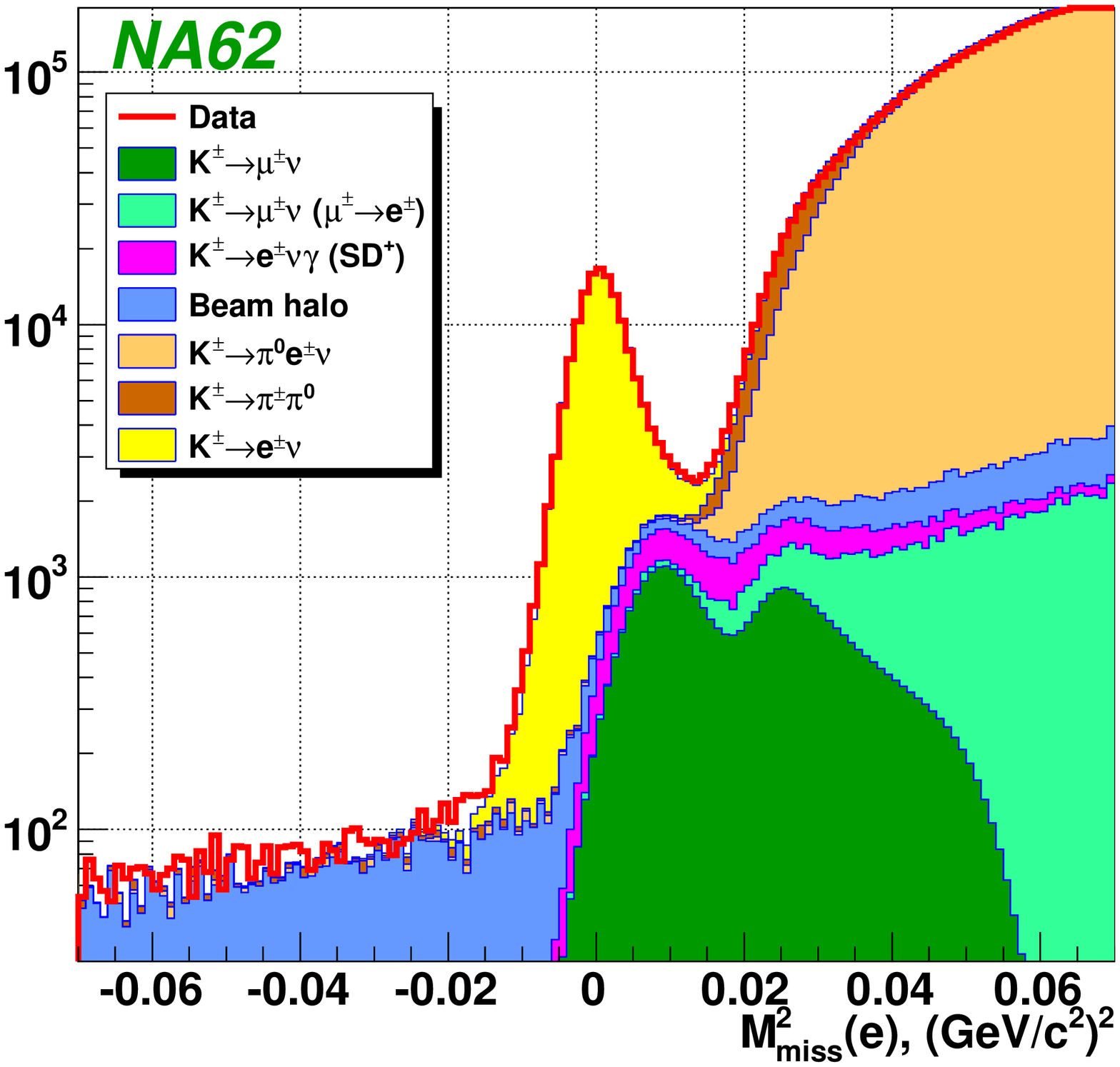}}%
\resizebox{0.4\textwidth}{!}{\includegraphics{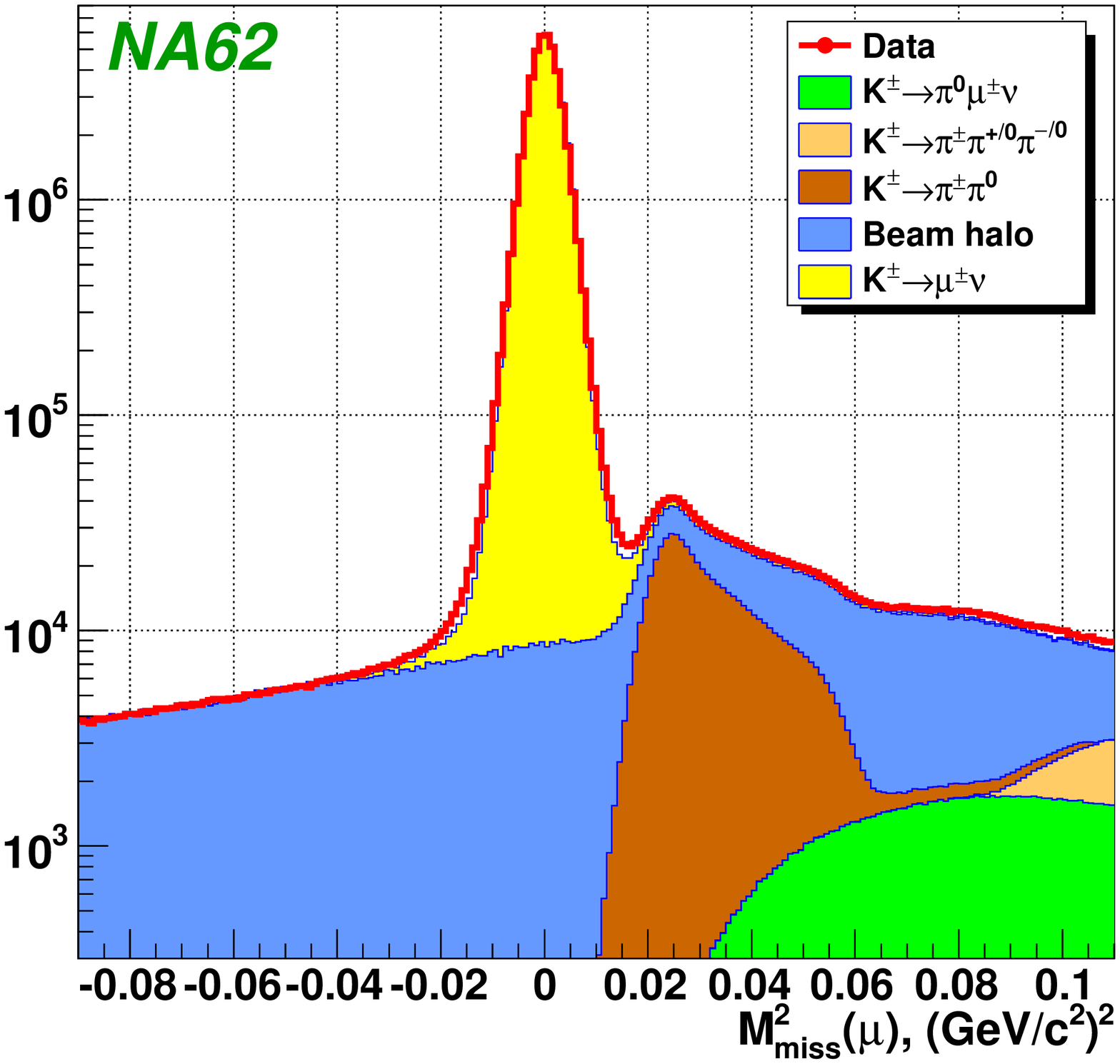}}%
\end{center}
\vspace{-6mm} \caption{Distributions of the reconstructed squared
missing mass $M_{\mathrm{miss}}^2(e)$ (left) and
$M_{\mathrm{miss}}^2(\mu)$ (right) of the $K_{e2}$ ($K_{\mu 2}$)
candidates compared with the sums of normalised estimated signal and
background components.} \label{fig:mm2}
\end{figure}

The analysis strategy is based on counting the numbers of
reconstructed $K_{e2}$ and $K_{\mu 2}$ candidates collected
concurrently. Therefore the analysis does not rely on the absolute
beam flux measurement, and several systematic effects cancel at
first order. The study is performed independently for 40 data
samples (10 bins of reconstructed lepton momentum and 4 samples with
different data taking conditions) by computing the ratio $R_K$ as
\begin{equation}
R_K = \frac{1}{D}\cdot \frac{N(K_{e2})-N_{\rm
B}(K_{e2})}{N(K_{\mu2}) - N_{\rm B}(K_{\mu2})}\cdot
\frac{A(K_{\mu2})}{A(K_{e2})} \cdot
\frac{f_\mu\times\epsilon(K_{\mu2})}
{f_e\times\epsilon(K_{e2})}\cdot\frac{1}{f_\mathrm{LKr}},
\label{eq:rkcomp}
\end{equation}
where $N(K_{\ell 2})$ are the numbers of selected $K_{\ell 2}$ candidates $(\ell=e,\mu)$, $N_{\rm B}(K_{\ell 2})$ are the numbers of background events, $A(K_{\mu 2})/A(K_{e2})$ is the geometric acceptance correction, $f_\ell$ are the efficiencies of $e$/$\mu$ identification, $\epsilon(K_{\ell 2})$ are the trigger efficiencies, $f_\mathrm{LKr}$ is the global efficiency of the LKr calorimeter
readout (which provides the information used for electron identification), and $D$ is the downscaling factor of the $K_{\mu2}$ trigger. The data sample is characterized by high values of $f_\ell$ and $\epsilon(K_{\ell 2})$ well above 99\%. A Monte Carlo (MC) simulation is used to evaluate the acceptance correction and the geometric part of the acceptances for most background processes entering the computation of $N_B(K_{\ell 2})$. Particle identification, trigger and readout efficiencies and the beam halo background are measured directly from control data samples.

Two selection criteria are used to distinguish $K_{e2}$ and $K_{\mu2}$ decays. Kinematic identification is based on the reconstructed squared missing mass assuming the track to be a electron or a muon: $M_{\mathrm{miss}}^2(\ell) = (P_K - P_\ell)^2$, where $P_K$ and $P_\ell$ ($\ell = e,\mu$) are the kaon and lepton
4-momenta (Fig.~\ref{fig:mm2}). A selection condition $M_1^2<M_{\mathrm{miss}}^2(\ell)<M_2^2$ is applied; $M_{1,2}^2$ vary across the lepton momentum bins depending on resolution. Lepton type identification is based on the ratio $E/p$ of energy deposit in the LKr calorimeter to track momentum measured by the spectrometer. Particles with $(E/p)_{\rm min}<E/p<1.1$ ($E/p<0.85$) are identified as electrons (muons), where $(E/p)_{\rm min}$ is 0.90 or 0.95,
depending on momentum.

\begin{figure}[tb]
\begin{center}
\vspace{-2mm}
\resizebox{0.48\textwidth}{!}{\includegraphics{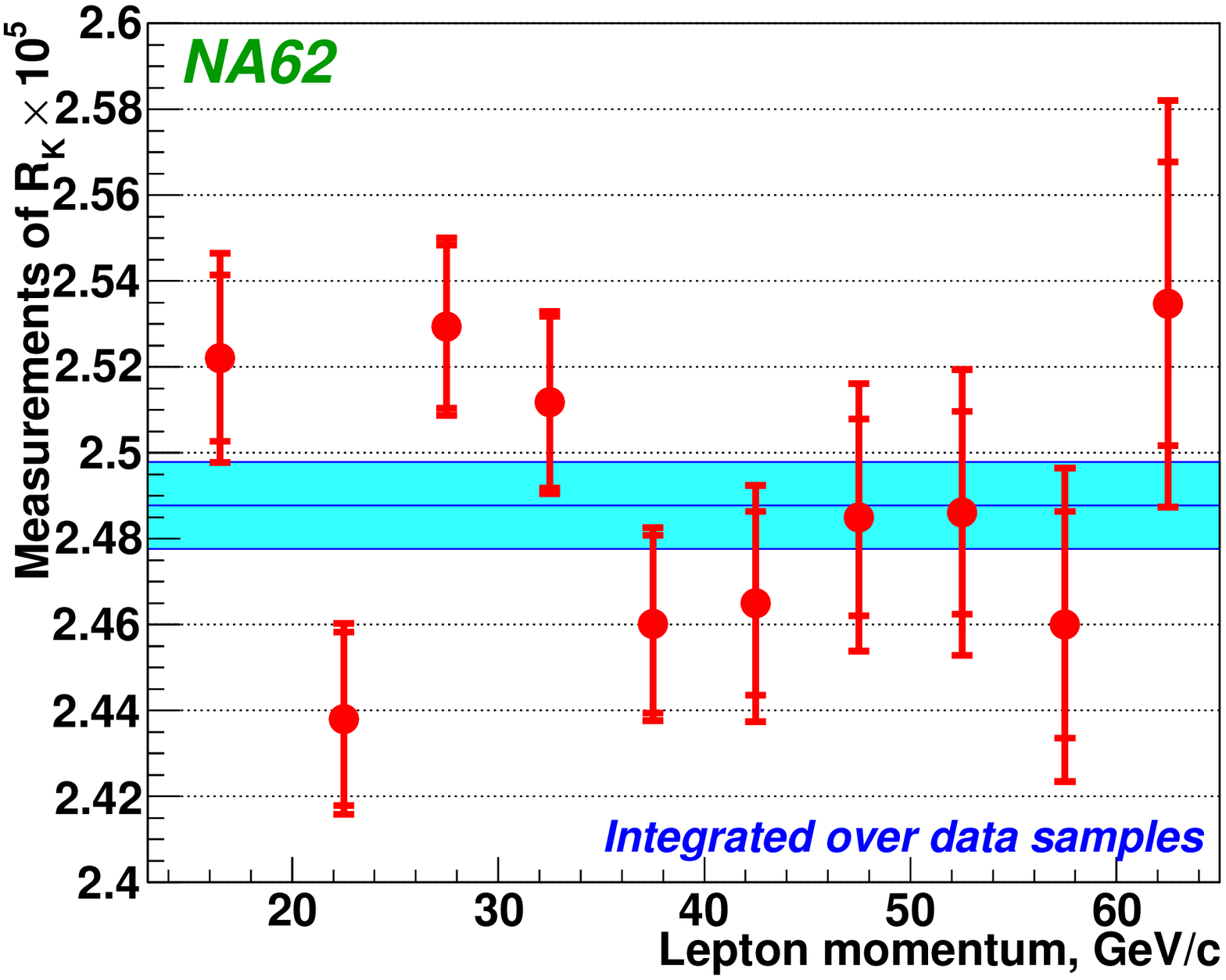}}%
\resizebox{0.32\textwidth}{!}{\includegraphics{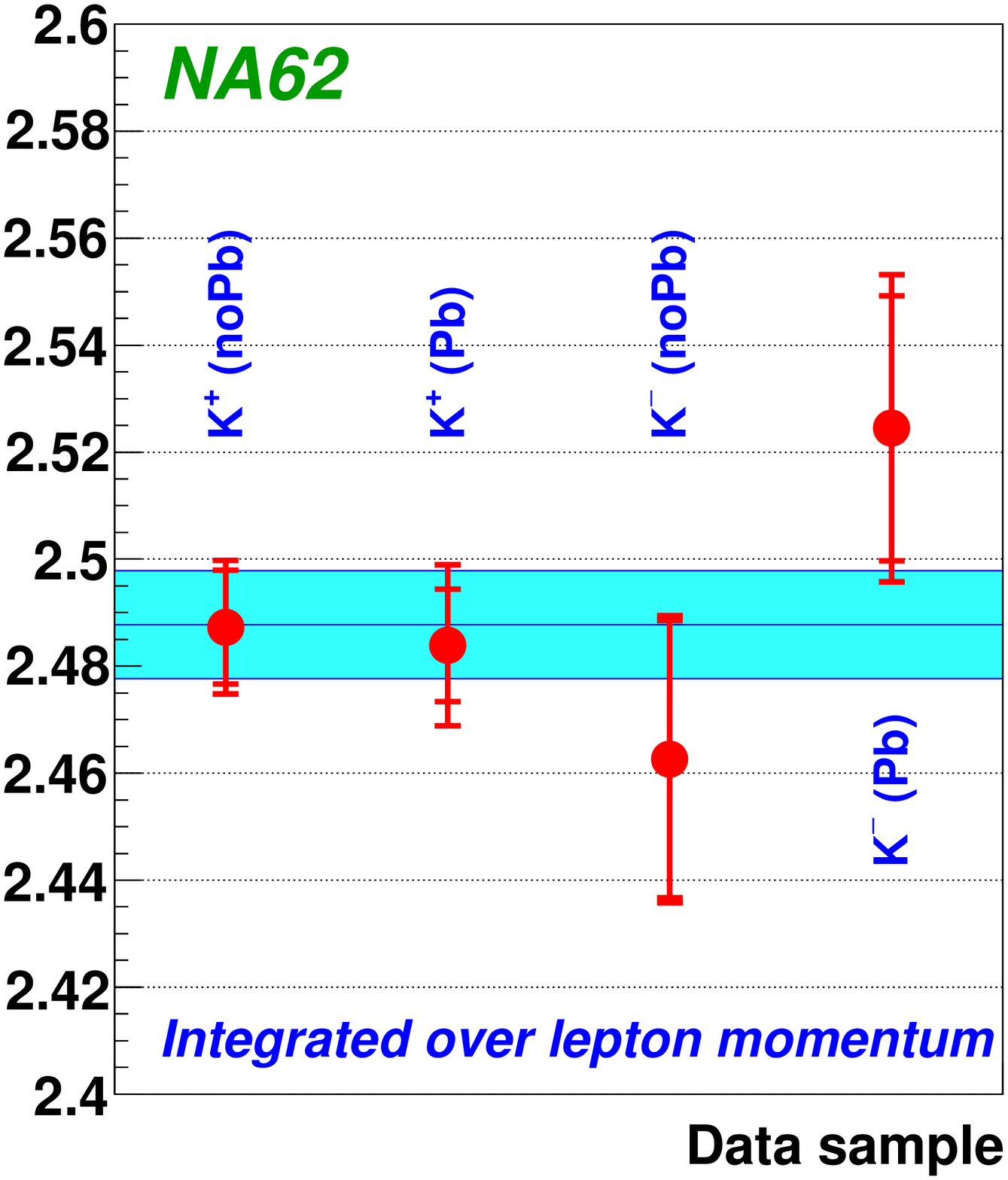}}%
\end{center}
\vspace{-6mm} \caption{Stability of the $R_K$ measurement versus
lepton momentum and for independent data samples. The ``Pb/noPb''
labels indicate samples with present and absent lead wall covering a
part of the LKr calorimeter in order to measure the muon
mis-identification probability. The result of the combined fit over
the 40 data bins is shown by horizontal bands.}\label{fig:rkfit}
\end{figure}

The numbers of selected $K_{e2}$ and $K_{\mu 2}$ candidates are 145,958 and $4.2817\times 10^7$ (the latter pre-scaled at trigger level). The background contamination in the $K_{e2}$ sample has been estimated by MC simulations and, where possible, direct measurements to be $(10.95\pm0.27)\%$. The largest background contribution is the $K_{\mu2}$ decay with a mis-identified muon via the `catastrophic' bremsstrahlung process in the LKr. To reduce the uncertainty due to background subtraction, the muon mis-identification probability $P_{\mu e}$ has been measured as a function of momentum using dedicated data samples. The contributions to the systematic uncertainty of the result include the uncertainties on the backgrounds, helium purity in the spectrometer tank (which influences the detection efficiency via bremsstrahlung and scattering), beam simulation, spectrometer alignment, particle identification and trigger efficiency. The result of the measurement, combined over the 40 independent samples taking into account correlations between the systematic errors, is
\begin{equation}
R_K = (2.488\pm 0.007_{\mathrm{stat.}}\pm 0.007_{\mathrm{syst.}})\times 10^{-5} = (2.488\pm0.010)\times 10^{-5}.
\end{equation}
The stability of $R_K$ measurements in lepton momentum bins and for
the separate data samples is shown in Fig.~\ref{fig:rkfit}. The
result is consistent with the Standard Model expectation, and the
achieved precision dominates the world average.

\boldmath
\section{$K^\pm\to\pi^\pm\gamma\gamma$ DECAY MEASUREMENTS WITH 2004 AND 2007 DATA}
\unboldmath

\begin{figure}[tb]
\begin{center}
\vspace{-2mm}
\resizebox{0.4\textwidth}{!}{\includegraphics{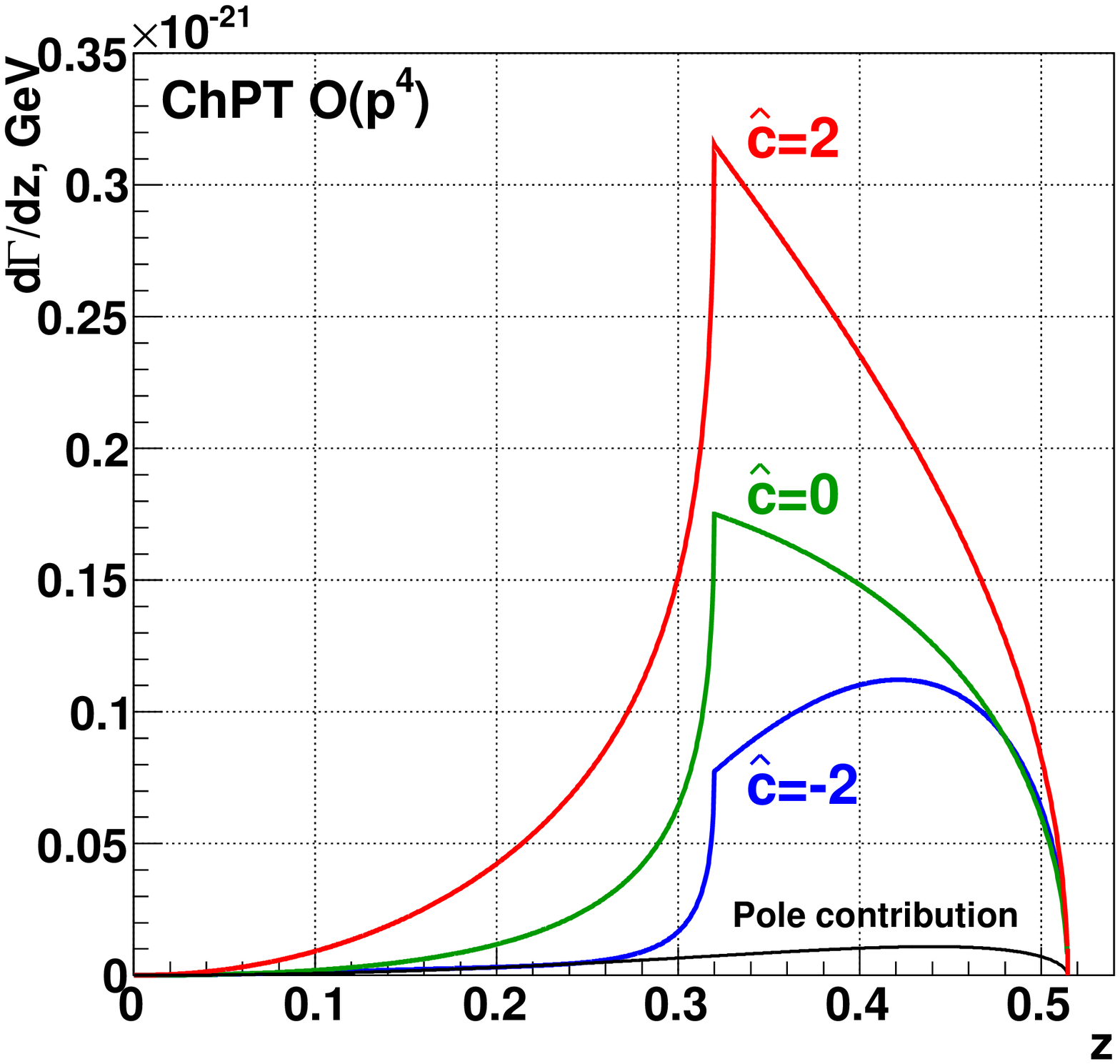}}%
\resizebox{0.4\textwidth}{!}{\includegraphics{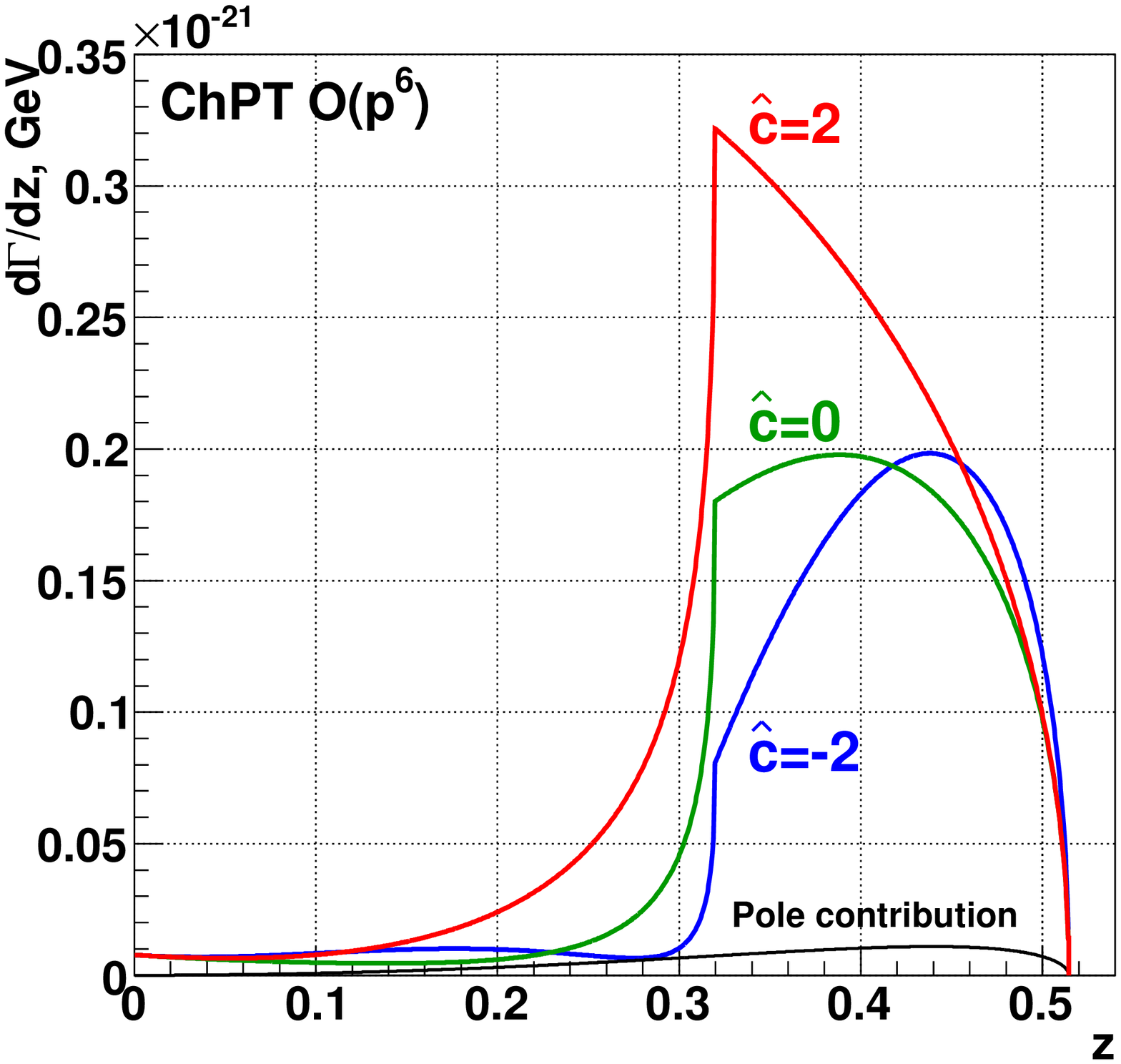}}%
\end{center}
\vspace{-9mm} \caption{ChPT predictions for the $K^\pm\to\pi^\pm\gamma\gamma$ differential rate in terms of $z=(m_{\gamma\gamma}/m_K)^2$ for ${\cal O}(p^4)$ and ${\cal O}(p^6)$ parameterizations according to~\cite{da96}, with $\hat{c}=-2; 0; 2$. The $\hat{c}$-independent pole contribution is also shown. For the ${\cal O}(p^6)$ parameterization, values of polynomial contributions~\cite{da96} $\eta_i=0$ and $K^\pm\to3\pi^\pm$ amplitude parameters from a fit to experimental data~\cite{bi03} are used.}\label{fig:br-vs-c}
\end{figure}

Measurements of radiative non-leptonic kaon decays provide crucial tests for the ability of the Chiral Perturbation Theory (ChPT) to explain weak low energy processes. In the ChPT framework, the $K^\pm\to\pi^\pm\gamma\gamma$ decay receives two non-interfering contributions at lowest non-trivial order ${\cal O}(p^4)$: the pion and kaon {\it loop amplitude} depending on an unknown ${\cal O}(1)$ constant $\hat{c}$ representing the total contribution of the counterterms, and the {\it pole amplitude}~\cite{ec88}. Higher order unitarity corrections from $K\to3\pi$ decays, including the main ${\cal O}(p^6)$ contribution as well as those beyond ${\cal O}(p^6)$ due to using the phenomenological values of $K\to3\pi$ amplitudes, have been found to modify the decay spectrum significantly; in particular, they lead to non-zero differential decay rate at zero diphoton invariant mass~\cite{da96}. The total decay rate is predicted to be ${\rm BR}(K^\pm\to\pi^\pm\gamma\gamma) \sim 10^{-6}$, with the pole amplitude contributing 5\% or less~\cite{da96, ge05}. The ChPT predictions for the decay spectra for several values of $\hat{c}$ are presented in Fig.~\ref{fig:br-vs-c}: the diphoton mass spectra exhibit a characteristic cusp at twice the pion mass due to the dominant pion loop amplitude. Experimentally, the only published $K^\pm\to\pi^\pm\gamma\gamma$ observation is that of 31 $K^+$ decay candidates in the kinematic region $100~{\rm MeV}/c<p_\pi^*<180~{\rm MeV}/c$ ($p_\pi^*$ is the $\pi^+$ momentum in the $K^+$ frame) by the BNL E787 experiment~\cite{ki97}.

\begin{figure}[tb]
\begin{center}
\vspace{-2mm}
\resizebox{0.4\textwidth}{!}{\includegraphics{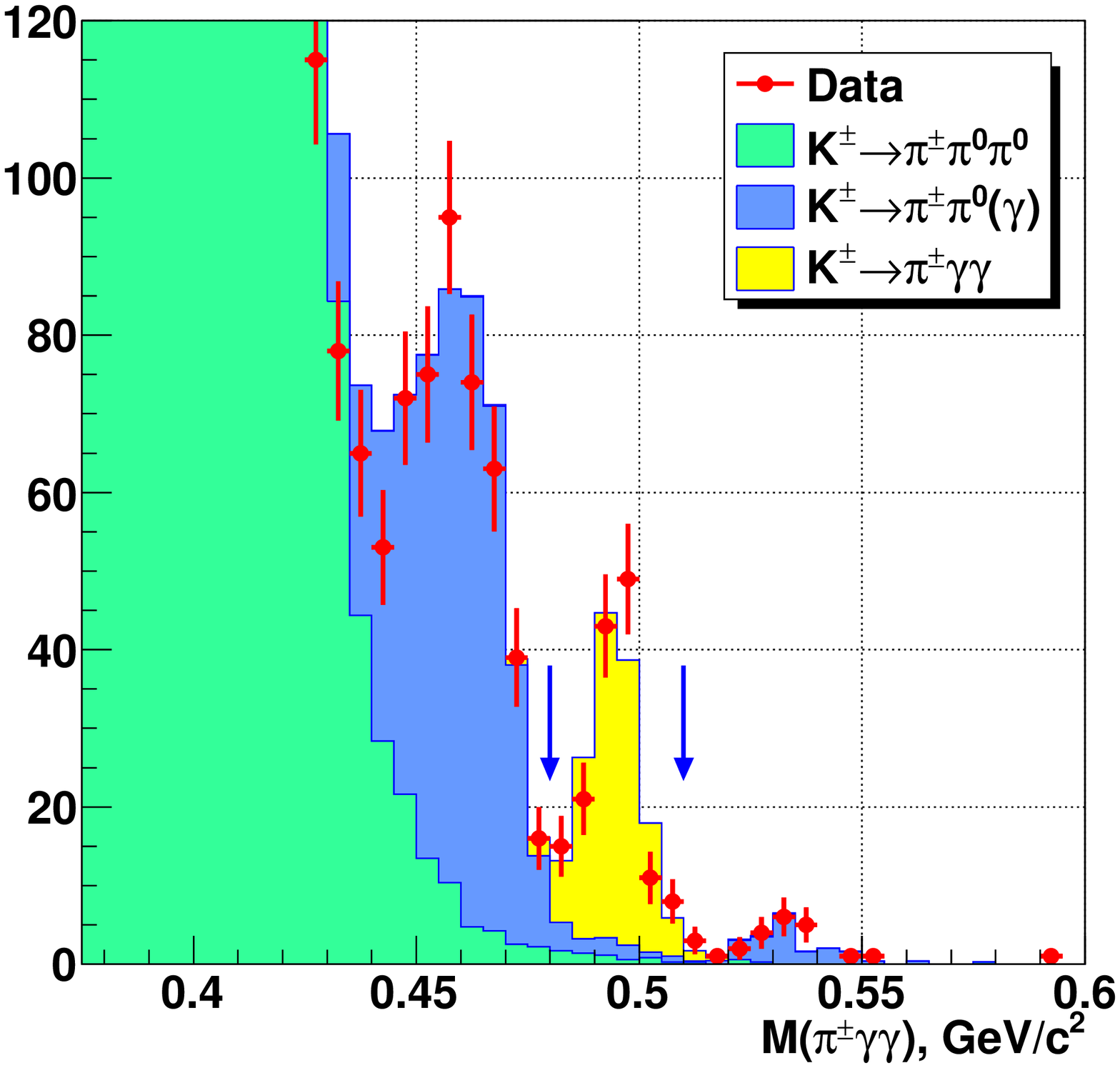}}%
\resizebox{0.4\textwidth}{!}{\includegraphics{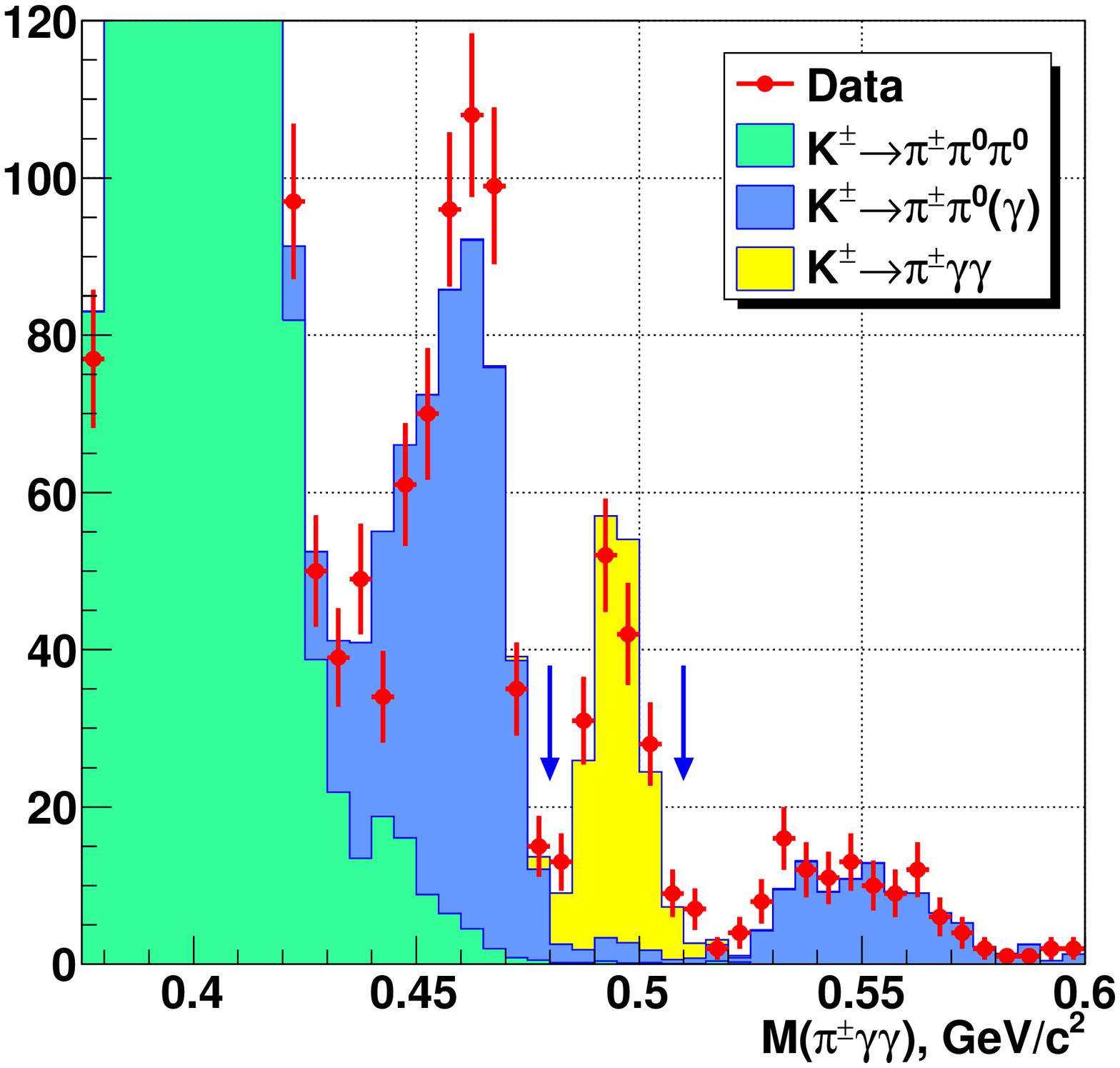}}%
\end{center}
\vspace{-8mm} \caption{The spectra of $\pi^\pm\gamma\gamma$ invariant mass with MC expectations for signal and backgrounds: 2004 data (left) and 2007 data (right). The signal region is indicated with arrows.} \label{fig:pigg-m}
\end{figure}

New measurements of this decay have been performed using minimum bias data sets collected during a 3-day special NA48/2 run in 2004 with 60~GeV/$c$ $K^\pm$ beams, and a 3-month NA62 run in 2007 with 74~GeV/$c$ $K^\pm$ beams. The latter set has been collected with a set of downscaled trigger conditions with an effective downscaling factor of about 20. The effective kaon fluxes collected in 2004 and 2007 are similar, but the background conditions and resolution on kinematic variables differ significantly.


Signal events are selected in the region $z=(m_{\gamma\gamma}/m_K)^2>0.2$ to reject the $K^\pm\to\pi^\pm\pi^0$ background peaking at $z=0.075$. The $\pi^\pm\gamma\gamma$ mass spectra, with the MC simulation expectations of the signal and background contributions, are displayed in Fig.~\ref{fig:pigg-m}: 147 (175) decays candidates are observed in the 2004 (2007) data set, with backgrounds contaminations of 12\% (7\%) from $K^\pm\to\pi^\pm\pi^0(\pi^0)(\gamma)$ decays with merged photon clusters in the LKr calorimeter.

\begin{figure}[tb]
\begin{center}
\vspace{-2mm}
\resizebox{0.4\textwidth}{!}{\includegraphics{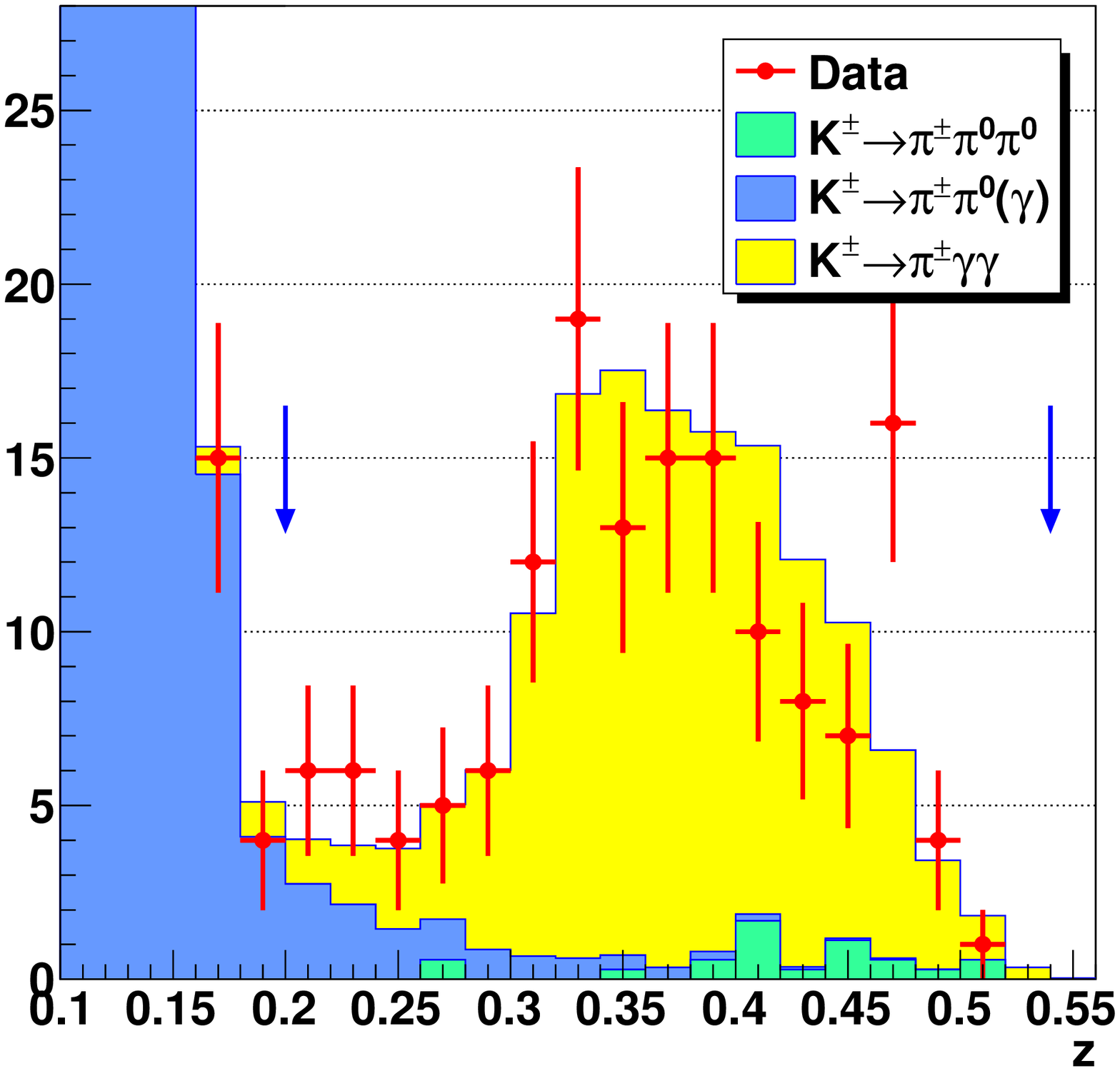}}%
\resizebox{0.4\textwidth}{!}{\includegraphics{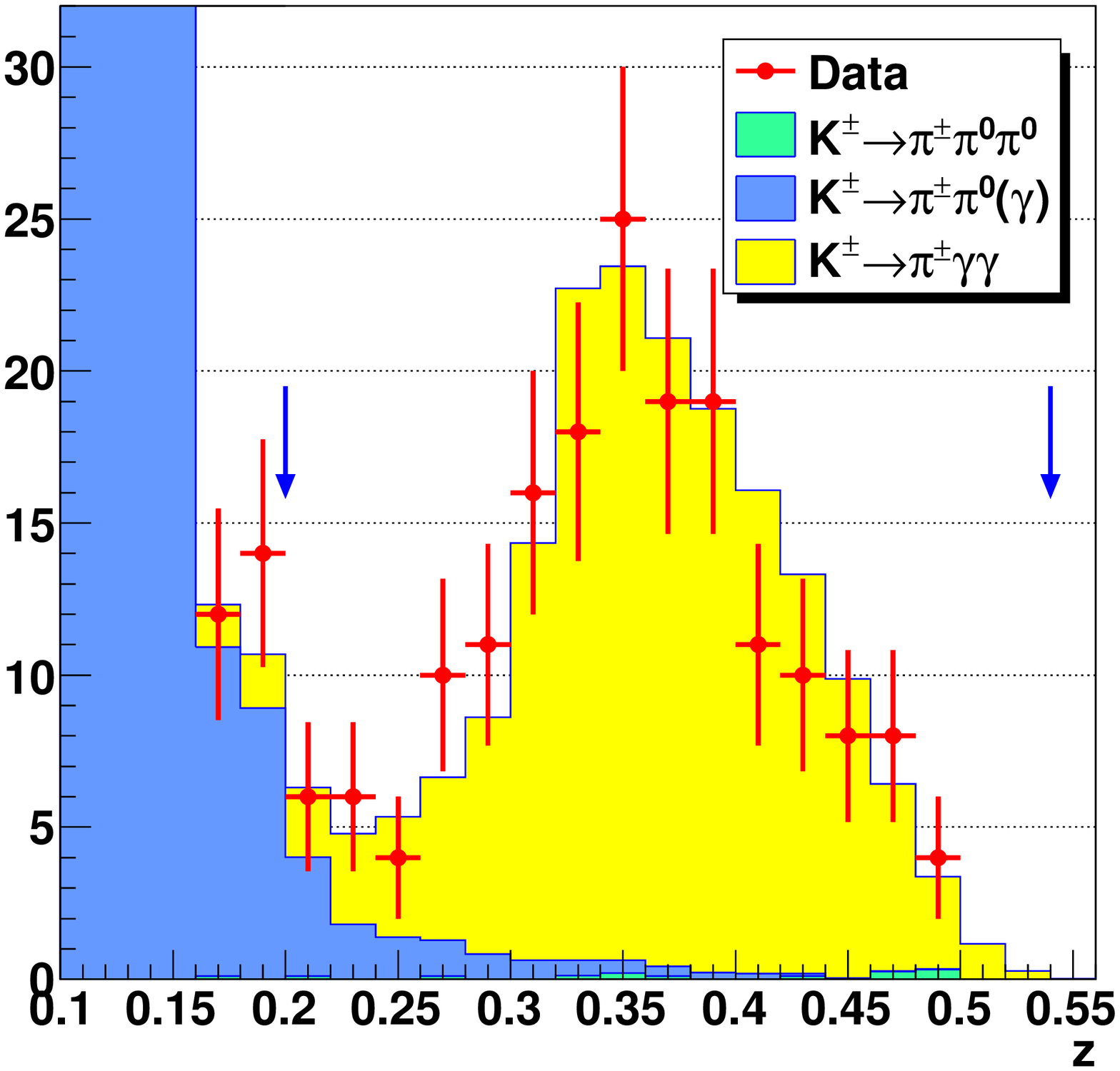}}%
\end{center}
\vspace{-8mm} \caption{The spectra of $z=(m_{\gamma\gamma}/m_K)^2$ with MC expectations for signal (best fit) and backgrounds: 2004 data (left) and 2007 data (right). The signal region ($0.2<z<0.52$) is indicated with arrows.} \vspace{-4mm} \label{fig:pigg-z}
\end{figure}

The data spectra of the $z$ kinematic variable, together with the signal and background expectations, are displayed in Fig.~\ref{fig:pigg-z}: they clearly exhibit the cusp at two-pion threshold as predicted by the ChPT. The values of the $\hat{c}$ parameter in the framework of the ChPT ${\cal O}(p^4)$ and ${\cal O}(p^6)$ parameterizations according to~\cite{da96} have been measured by the performing likelihood fits to the data. The preliminary results of the fits are presented in Table~\ref{tab:chat}: they are in agreement with the earlier BNL E787 ones. The uncertainties are dominated by the statistical ones; the systematic errors and mainly due to uncertainties of the background estimates. The ${\cal O}(p^6)$ parametrization involves a number of external inputs. In this analysis, they have been fixed as follows: the polynomial contribution terms are $\eta_1=2.06$, $\eta_2=0.24$ and $\eta_3=-0.26$ as suggested in~\cite{da96}, while the $K^\pm\to3\pi^\pm$ amplitude parameters come from a fit to the experimental data~\cite{bi03}. Along with the separate 2004 and 2007 results, the combined results are presented in Table~\ref{tab:chat}. The combination takes into account the large positive correlation of the systematic uncertainties of the two measurements.

\begin{table}[tb]
\begin{center}
\caption{The preliminary results of the fits to the $K^\pm\to\pi^\pm\gamma\gamma$ diphoton mass spectra to the ChPT parameterizations~\cite{da96}. The quoted BR values correspond to the full kinematic range.}
\label{tab:chat}
\begin{tabular}{l|ccc}
\hline & 2004 data & 2007 data & Combined\\
\hline
$\hat{c}$, ${\cal O}(p^4)$ fit &
$1.36\pm0.33_{\rm stat}\pm0.07_{\rm syst}$ &
$1.71\pm0.29_{\rm stat}\pm0.06_{\rm syst}$ &
$1.56\pm0.22_{\rm stat}\pm0.07_{\rm syst}$\\
$\hat{c}$, ${\cal O}(p^6)$ fit &
$1.67\pm0.39_{\rm stat}\pm0.09_{\rm syst}$ &
$2.21\pm0.31_{\rm stat}\pm0.08_{\rm syst}$ &
$2.00\pm0.24_{\rm stat}\pm0.09_{\rm syst}$\\
BR, ${\cal O}(p^6)$ fit        &
$(0.94\pm0.08)\times10^{-6}$ &
$(1.06\pm0.07)\times10^{-6}$ &
$(1.01\pm0.06)\times 10^{-6}$\\
\hline
\end{tabular}
\end{center}
\vspace{-2mm}
\end{table}

\boldmath
\section{THE ULTRA-RARE DECAY $K^+\to\pi^+\nu\bar\nu$}
\unboldmath

Among the flavour changing neutral current $K$ and $B$ decays, the  $K\to\pi\nu\bar\nu$ decays play a key role in the search for new physics through the underlying mechanisms of flavour mixing. These decays are strongly suppressed in the SM (the highest CKM suppression), and are dominated by top-quark loop contributions. The SM branching ratios have been computed to an exceptionally high
precision with respect to other loop-induced meson decays: ${\rm
BR}(K^+\to\pi^+\nu\bar\nu)=8.22(75)\times 10^{-11}$ and ${\rm
BR}(K_L\to\pi^0\nu\bar\nu)=2.57(37)\times 10^{-11}$; the uncertainties are dominated by parametric ones, and the irreducible theoretical uncertainties are at a $\sim 1\%$ level~\cite{br11}. The extreme theoretical cleanness of these decays remains also in certain new physics scenarios. Experimentally, the $K^+\to\pi^+\nu\bar\nu$ decay has been observed by the BNL E787/E949 experiments, and the measured branching ratio is
$\left(1.73^{+1.15}_{-1.05}\right)\times 10^{-10}$~\cite{ar09}. The
achieved precision is inferior to that of the SM expectation.

The main goal of the NA62 experiment at CERN is the measurement of the $K^+\to\pi^+\nu\bar\nu$ decay rate at the 10\% precision level, which would constitute a significant test of the SM. The experiment is expected to collect about 100 signal events in two years of data taking, keeping the systematic uncertainties and backgrounds low. Assuming a 10\% signal acceptance and the SM decay rate, the kaon flux should correspond to at least $10^{13}$ $K^+$ decays in the fiducial volume. In order to achieve a small systematic uncertainty, a rejection factor for generic kaon decays of the order of $10^{12}$ is required, and the background suppression factors need to be measured directly from the data. In order to achieve the required kaon intensity, signal acceptance and
background suppression, most of the NA48/NA62 apparatus used until 2008
is being replaced with new detectors. The CERN SPS extraction line used by the NA48 experiment is capable of delivering beam intensity sufficient for the NA62. Consequently the new setup is housed at the CERN North Area High Intensity Facility where the NA48 was located. The decay in flight technique will be used; optimisation of the signal acceptance drives the
choice of a 75 GeV/$c$ charged kaon beam with 1\% momentum bite. The
experimental setup is conceptually similar to the one used for NA48:
a $\sim 100$~m long beam line to form the appropriate secondary
beam, a $\sim 80$~m long evacuated decay volume, and a series of
downstream detectors measuring the secondary particles from the
$K^+$ decays in the fiducial decay volume.

The signal signature is one track in the final state matched to one $K^+$ track in the beam. The integrated rate upstream is about 800 MHz (only 6\% of the beam particles are kaons, the others being mostly $\pi^+$ and protons). The rate seen by the detector downstream is about 10 MHz, mainly due to $K^+$ decays. Timing and
spatial information are required to match the upstream and downstream track. Backgrounds come from kaon decays with a single reconstructed track in the final state, including accidentally matched upstream and downstream tracks. The background suppression profits from the high kaon beam momentum. A variety of techniques will be employed in combination in order to reach the required level of background rejection. They can be schematically divided into kinematic rejection, precise timing, highly efficient photon and muon veto systems, and precise particle identification systems to distinguish $\pi^+$, $K^+$ and positrons. The above requirements drove the design and the construction of the subdetector systems.

The main NA62 subdetectors are: a differential Cherenkov counter (CEDAR) on the beam line to identify the $K^+$ in the beam; a silicon pixel beam tracker; guard-ring counters surrounding the beam tracker to veto catastrophic interactions of particles; a downstream spectrometer composed of 4 straw chambers operating in vacuum; a RICH detector to distinguish pions and muons; a scintillator hodoscope; a muon veto detector. The photon veto detectors will include a series of annular lead glass calorimeters surrounding the decay and detector volume, the NA48 LKr calorimeter, and two small angle calorimeters to provide hermetic coverage for photons emitted at close to zero angle to the beam. The design of the experimental apparatus and the R\&D of the new subdetectors have been completed. The experiment is under construction, and the first technical run is scheduled for October--December 2012.

\section*{CONCLUSIONS}

The NA48/2 and NA62 experiments at CERN have recently accomplished several precision measurements of rare $K^\pm$ decays. The $R_K$ phase of the NA62 experiment provided the most precise measurement of the ratio of leptonic $K^\pm$ decay rates $R_K=(2.488\pm0.010)\times 10^{-5}$. This result is consistent with the SM expectation, and constrains multi-Higgs and fourth generation new physics scenarios. NA48/2 and NA62 experiments have performed new measurements of the $K^\pm\to\pi^\pm\gamma\gamma$ decay, significantly improving the precision of ChPT tests with this channel. The ultra-rare $K^+\to\pi^+\nu\bar\nu$ decay represents a unique environment to search for new physics. The NA62 experiment, aiming to collect ${\cal O}(100)$ events of this decay, is being constructed and is
preparing for a technical run in 2012.


\end{document}